\documentclass[superscriptaddress,aps, prd, reprint, nofootinbib]{revtex4-2}

\usepackage{amsmath, amssymb, amsbsy, amstext, amsthm,mathtools}
\usepackage[usenames,dvipsnames,svgnames,table]{xcolor}
\usepackage{hyperref}
\hypersetup{hidelinks}

\DeclareMathOperator{\tr}{tr}
\DeclareMathOperator{\Path}{P}

\newcommand{\ud}{\mathrm{d}}
\newcommand{\lab}[1]{{\mathrm{#1}}}
\newcommand{\slab}[1]{{\textsc{#1}}}
\newcommand{\mb}[1]{{\mathbf{#1}}}
\newcommand{\minus}{{\scalebox {0.75}[1.0]{$-$}}}
\newcommand{\sminus}{{\scalebox {0.5}[0.7]{$-$}}}
\newcommand{\labM}{{\textsc{m}}}
\newcommand{\labD}{{\scalebox{0.65}{$\Delta$}}}
\newcommand{\labW}{{\textsc{w}}}
\newcommand{\es}{\hspace{0.5pt}}
\newcommand{\wG}{g}

\definecolor{cobalt}{RGB}{44, 98, 120}

\begin{document}

		\title{Axion Mass from Magnetic Monopole Loops}
		\date{\today}

		\author{JiJi Fan}
		\affiliation{Department of Physics \& Brown Theoretical Physics Center, Brown University, Providence, RI, 02912, USA}

		\author{Katherine Fraser}
		\affiliation{Department of Physics, Harvard University, Cambridge, MA, 02138, USA}

		\author{Matthew Reece}
		\affiliation{Department of Physics, Harvard University, Cambridge, MA, 02138, USA}

		\author{John Stout}
		\affiliation{Department of Physics, Harvard University, Cambridge, MA, 02138, USA}

		\begin{abstract}
			 We show that axions interacting with abelian gauge fields obtain a potential from loops of magnetic monopoles. This is a consequence of the Witten effect: the axion field causes the monopoles to acquire an electric charge and alters their energy spectrum. The axion potential can also be understood as a type of instanton effect due to a Euclidean monopole worldline winding around its dyon collective coordinate. We calculate this effect, which has features in common with both nonabelian instantons and Euclidean brane instantons. To provide consistency checks, we argue that this axion potential vanishes in the presence of a massless charged fermion and that it is robust against the presence of higher-derivative corrections in the effective Lagrangian. Finally, as a first step toward connecting with particle phenomenology and cosmology, we discuss the regime in which this potential is important in determining the dark matter relic abundance in a hidden sector containing an abelian gauge group, monopoles, and axions.
		\end{abstract}

		\maketitle

		\section{Introduction}
		 
		Axions are naturally light scalar bosons that are of great interest in solving the strong CP problem~\cite{Peccei:1977hh, Peccei:1977ur,Weinberg:1977ma,Wilczek:1977pj}, as dark matter candidates~\cite{Preskill:1982cy,Dine:1982ah,Abbott:1982af}, and for many other applications. It is well known that instanton effects can generate a potential for an axion $\theta$~\cite{Weinberg:1977ma,Wilczek:1977pj} when it is coupled to a nonabelian gauge field via the topological coupling $\theta \tr(F \wedge F)$. Even in the absence of axion interactions with gauge fields, it is known that Euclidean branes can give rise to axion potentials~\cite{Dine:1986zy,Dine:1987bq,Becker:1995kb,Witten:1996bn,Ooguri:1996me}. Here, we argue that axions coupled to {\em abelian} gauge fields through a $\theta F \wedge F$ coupling acquire a potential through an instanton effect whenever there are monopoles magnetically charged under $F$, due to the Witten effect~\cite{Witten:1979ey}. Like nonabelian instantons, these effects are associated with 4d gauge theory dynamics. Like Euclidean branes, they occur within a well-behaved semiclassical expansion free of infrared divergences. In fact, we expect that our instantons are continuously connected to, or a limiting case of, known instanton effects in specific UV completions~\cite{Heidenreich:2020pkc}. The virtue of our approach is that, by working from the bottom up, we deduce that such effects must exist \cite{Stout:2020uaf} even when we do not know the UV theory.\footnote{The existence of the instantons we discuss here has been noted previously by Jake McNamara~\cite{McNamara:2020} and communicated to MR in the course of writing~\cite{Heidenreich:2020pkc}, although neither considered computing an axion potential from them at the time.} 
		
		The Completeness Hypothesis postulates that any UV-complete theory of an interacting $U(1)$ gauge field (which has quantized charge) contains magnetic monopoles~\cite{Polchinski:2003bq}, which break a would-be 1-form global symmetry~\cite{Gaiotto:2014kfa}. This is, in particular, believed to be true of all theories of quantum gravity~\cite{ArkaniHamed:2006dz,Banks:2010zn,delaFuente:2014aca,Harlow:2018tng}. Assuming the validity of the Completeness Hypothesis, the instanton effect that we discuss will give rise to an effective potential for any axion interacting with photons. This is of great phenomenological interest, since the $\theta F \wedge F$ interaction is the primary target of experimental searches for axions~\cite{Sikivie:1983ip, Sikivie:1985yu, Asztalos:2009yp, Graham:2015ouw}.

		We consider an effective theory of a periodic axion field $\theta \cong \theta + 2\pi$ coupled to a gauge field $A$ normalized such that the allowed Wilson lines $\Path[\es\exp(i \es q  \!\oint\! A)\es]$ have integer charge $q \in \mathbb{Z}$:
		\begin{equation}  \label{eq:Sbulk}
		S = \int   \left[\frac{1}{2} f^2 \es \ud \theta \wedge \star \ud \theta - \frac{1}{2e^2} F \wedge  \star F + \frac{k \es \theta}{8\pi^2} F \wedge F\right].		
		\end{equation}
		The axion-gauge field coupling is of Chern-Simons type, with quantized coefficient $k \in \mathbb{Z}$. Through the Witten effect, a magnetic monopole in the presence of a nonzero background $\theta$ acquires an electric charge $-k\es\theta/(2\pi)$. A consistent description of this effect requires that the effective theory on the magnetic monopole worldvolume contains, in addition to the usual translational zero modes $x^\mu$, a collective coordinate interacting with the field $\theta$. This takes the form of a compact scalar boson $\sigma \cong \sigma + 2\pi$, with an action that (expanding around a monopole worldline extended in time) contains~\cite{Jackiw:1975ep}:
		\begin{equation}  \label{eq:Smono}
		S = \int_\gamma \left[\frac{1}{2} l_\sigma \ud_A \sigma \wedge \star \ud_A \sigma + \frac{\theta}{2\pi} \ud_A \sigma\right],
		\end{equation}
		where the gauge-covariant derivative $\ud_A \sigma \equiv \ud \sigma + k A$ respects a shift of $\sigma$ under $A$ gauge transformations. The mode $\sigma$ behaves as a quantum particle on a circle (see, e.g., App.~D.1 of \cite{Gaiotto:2017yup}). Its energy eigenstates, labeled by integers $n \in \mathbb{Z}$, correspond to dyonic states of the monopole with electric charge $k\left(n - {\theta}/{2\pi}\right)$ and energy 
		\begin{equation}  \label{eq:dyonenergy}
		E_n = \frac{1}{2 l_\sigma} \!\left(n - \frac{\theta}{2\pi}\right)^2.
		\end{equation}
		There is a monodromy $n \mapsto n+1$ when $\theta \mapsto \theta+2\pi$ that ensures the spectrum of the theory is periodic.

		 We can estimate $l_\sigma$ by comparing \eqref{eq:dyonenergy} to the energy of the classical field configuration outside a monopole in an axion background, following~\cite{Fischler:1983sc}, from which we obtain: \begin{samepage}
		 \begin{equation} \label{eq:matching}
		 l_\sigma \sim \frac{4\pi}{e^2 k^2} r_*\es, \quad r_* = \max(r_c, r_0)\es,
		 \end{equation}
		 where $r_c = \pi/(e^2 m_\slab{m})$\end{samepage} is the classical radius of the  magnetic monopole (of mass $m_\labM$) and $r_0 = {ke}/{(8\pi^2 f)}$ is the length scale over which the axion field is screened near the monopole core. In the special case of critical 't~Hooft-Polyakov monopoles~\cite{hooft:1974qc,polyakov:1974ek}, we begin with an $SU(2)$ gauge theory with coupling $\wG$. Matching to~\eqref{eq:Sbulk} gives $e = \wG/2$ and $k = 2$, while matching to~\eqref{eq:Smono} (when $r_c \gg r_0$) gives $l_\sigma = m_\labM/m_\labW^2$ where $m_\labM = 4\pi v/\wG$ and $m_\labW = \wG v$ is the W boson mass. (We have chosen the order-one coefficient in \eqref{eq:matching} to be accurate for this case, but it will differ in general theories.)
		 
		 Because the dyon energy spectrum \eqref{eq:dyonenergy} is $\theta$-dependent, we can integrate out the dyons and obtain an effective potential for $\theta$. This can be understood either as a sum of Coleman-Weinberg-type potentials~\cite{Coleman:1973jx} from each dyon mode $n$, or as a sum over loops with nontrivial winding of $\sigma$ around the loop. These two calculations are related by Poisson resummation, as we explain below. Although there is prior work on the $\theta$ potential generated by a gas of (non-virtual) monopoles and antimonopoles (see~\cite{Fischler:1983sc, Kawasaki:2015lpf, Nomura:2015xil, Kawasaki:2017xwt} and follow-ups), the effect of monopole loops on the vacuum $\theta$ potential is, as far as we know, absent from the prior literature.

		\section{Monopole Loops}
			We would like to compute the vacuum energy in the presence of ``fundamental'' magnetic monopoles. Schematically, the vacuum energy should be derived by computing a Euclidean path integral of the form
			\begin{equation}
				\mathcal{Z}(\theta) = \sum_{\lab{worldlines}} \int\!\mathcal{D}(\lab{fields}) \, \lab{e}^{-S_\textsc{e}[\lab{fields}, \lab{worldlines}, \theta]} \,,
			\end{equation}
			and taking the limit of infinite spacetime volume,
			\begin{equation}
				V_\lab{eff}(\theta) = -\lim_{\mathcal{V} \to \infty} \frac{1}{\mathcal{V}} \log \mathcal{Z}(\theta)\,.
			\end{equation}
			The worldline formalism has previously been applied to other physical processes involving monopoles, e.g., to pair production in magnetic fields~\cite{Affleck:1981ag}.
			
			In the limit where interactions between the configurations are small, we expect the partition function to be dominated by disconnected vacuum paths characterized by the transition amplitude $Z_{\lab{S}^1}(\theta)$, the Feynman-weighted sum over all paths that are topologically a circle $\lab{S}^1$. 
			These contributions exponentiate:\begin{samepage}
			\begin{equation}
				\mathcal{Z}(\theta) = \sum_{n = 0}^{\infty}\frac{1}{n!} \left(Z_{\lab{S}^1}\right)^n = \exp\!\big(Z_{\lab{S}^1}\!\es(\theta)\big)\,.
			\end{equation}
		 	Hence\end{samepage} $V_\lab{eff}(\theta) = \minus \frac{1}{\mathcal{V}} Z_{\lab{S}^1}\!\es(\theta)$; we work in the first-quantized picture to compute the amplitude $Z_{\lab{S}^1}\es\!(\theta)$~\cite{Polchinski:1998rq}. We sum over all trajectories that return to the same configuration. This includes an integral over the invariant length (Schwinger proper time) $\tau$, weighted with a $1/2 \tau$ to account for overcounting trajectories related by translations and reflections. So,
			\begin{equation}
				Z_{\lab{S}^1} =  \int_{0}^{\infty} \! \frac{\ud \tau}{2 \tau}\,  Z(\tau, \theta)\,,
			\end{equation}
			with $Z(\tau, \theta)$ the sum over transition amplitudes at fixed $\theta$ of all trajectories with invariant length $\tau$.

There are two ways we can compute $Z_{\lab{S}^1}$. For a free particle of mass $m$, the gauge fixed transition amplitude for a trajectory of length $\tau$ from point $x$ to point $x'$ is
\begin{equation}
\langle x'|x \rangle_\tau = \frac{1}{2(2 \pi \tau)^{2}}\es\exp\!\left(-\frac{1}{2\tau}(x - x')^2 - m^2 \tau\right)
\end{equation}
After integrating over all trajectories that begin and end at the same point and canceling off the a factor of the spacetime volume from the measure with the factor in the definition of the effective potential, we obtain
\begin{equation}
\label{eqn:eff_pot}
V_\lab{eff} = - \int_0^\infty \!\frac{\ud\tau}{2 \tau} \frac{1}{2 (2 \pi \tau)^{2}}\es \exp\!\left(-\frac{m^2 \tau}{2}\right).
\end{equation}

We will sum over all dyon modes, labeled by $n \in \mathbb{Z}$. To simplify the computation, we assume that the dyon mass spectrum takes the form
\begin{equation} \label{eq:dyonmass}
m_n^2 = m_\labM^2 + m_\labD^2 \left(n - \frac{\theta}{2\pi}\right)^2, \quad m_\labD^2 = \frac{m_\labM}{l_\sigma}\,.
\end{equation}
This agrees with \eqref{eq:dyonenergy} to order $1/l_\sigma$, and in certain cases is an exact consequence of a BPS condition. In general, there may be power corrections in $(m_\labM l_\sigma)^{\sminus 1}$. Summing over the tower of states, we obtain the effective potential
\begin{equation}
-\sum_{n \in \mathbb{Z}} \int_0^\infty \!\!\!\es\frac{\ud\tau}{4\tau\,(2 \pi \tau)^{2}} \exp\!\left(\!\es-\frac{m_\labM^2 \tau}{2} - \frac{m_\labD^2 \tau}{2}\left(n - \frac{\theta}{2\pi}\right)^2\right).
\end{equation}
Periodicity in $\theta$, arising from the sum over $n$, is manifest after Poisson resummation:
			\begin{equation}
				\sum_{n \in \mathbb{Z}} \mathrm{e}^{\sminus\frac{1}{2}m_{\scalebox{0.5}{$\Delta$}}^2 \tau \left(n - \frac{\theta}{2 \pi}\right)^2} = \sum_{\ell \in \mathbb{Z}} \sqrt{\frac{2\pi}{m_\labD^2 \tau}} \exp\!\left(- \frac{2\pi^2\ell^2}{m_\labD^2 \tau} + i \ell \theta\right).
			\end{equation}
			The effective potential then becomes
			\begin{equation} 
				- \frac{\pi^2}{m_\labD} \sum_{\ell \in \mathbb{Z}} \int_{0}^{\infty}\!\!\frac{\ud \tau\, \lab{e}^{i \ell \theta}}{(2 \pi \tau)^{7/2}} \exp\!\left(- \frac{m_\labM^2 \tau}{2}  - \frac{2\pi^2\ell^2}{m_\labD^2 \tau}\right). \label{eq:effPot}
			\end{equation}
After integrating, the result is 
			\begin{align}  \label{eq:Veffresult}
				V_\lab{eff}(\theta) 
				&= -\sum_{\ell = 1}^\infty \frac{m_\labD^2 m_\labM^2}{32 \pi^4 \ell^3}  \mathrm{e}^{-2\pi \ell m_\labM/m_{\scalebox{0.5}{$\Delta$}}}  \cos(\ell \theta)\,\times && \nonumber \\
				&\qquad\qquad \left(1 + \frac{3 m_\labD}{2 \pi \ell m_\labM} + \frac{3 m_\labD^2}{(2\pi \ell m_\labM)^2}\right) ,
		        \end{align}
where we have ignored the irrelevant constant from the divergent  $\ell = 0$ integral.

\newpage 

We can think of the integer $\ell$ as the number of times the coordinate $\sigma$ winds around itself for a  particular configuration, and so we expect that we can interpret the effective potential (\ref{eq:effPot}) in terms of the monopole wordline action. 
Indeed, if we consider the relativistic completion of \eqref{eq:Smono} with the dyon collective coordinate $\sigma$ treated as another (compact) spatial direction in which the monopole 
\newpage \noindent
 propagates, analogous to the DBI action:
\begin{equation}
S_\labM = m_{\labM} \int_\gamma \ud \lambda\, \sqrt{\frac{\ud x_\mu}{\ud \lambda}\frac{\ud x^\mu}{\ud \lambda} + \frac{l_\sigma}{m_\labM}\! \left(\frac{\ud_A \sigma}{\ud \lambda}\right)^2} + \int_\gamma \frac{\theta}{2\pi} \ud_A \sigma\,,
\end{equation}
then we can compute the transition amplitude for a trajectory of length $\tau$ from point ($x,\sigma$) to point ($x',\sigma'$),
			\begin{widetext}

			\begin{equation}
				\langle x', \sigma' | x, \sigma \rangle_{\tau} =   \frac{1}{2(2 \pi \tau)^{5/2}} \es \exp\!\left(-\frac{1}{2\tau}(x' - x)^2 - \frac{l_\sigma}{2 m_\labM \tau}(\sigma' - \sigma)^2 - \frac{m_\labM^2 \tau}{2} + \frac{i \theta}{2 \pi} (\sigma' - \sigma)\right).
			\end{equation}
			
			\end{widetext}
Again, we integrate over all trajectories that begin and end at the same point, this time getting a contribution from the sum over windings $\sigma' - \sigma = 2 \pi \ell$, which nicely reproduces \eqref{eq:Veffresult}. 
					
			We can understand the exponential factor in~\eqref{eq:Veffresult} via a saddle point approximation for each $\ell$, corresponding to a classical Euclidean instanton solution that winds $\ell$ times in the $\sigma$ coordinate while remaining at constant $x^\mu$. The saddle is at Schwinger proper time $\tau_* = 2\pi\ell/(m_\labM m_\labD)$. The instanton action, which controls the convergence of the Fourier expansion (\ref{eq:Veffresult}), is
			\begin{equation}
				S = \frac{2\pi m_\labM}{m_\labD} \sim \frac{4\pi^2}{ke^2} \sqrt{\frac{\max(r_c, r_0)}{r_c}}.
			\end{equation}
			Remarkably, for the critical 't Hooft-Polyakov monopole, the instanton action is 
			$S = {8\pi^2}/{\wG^2}$,
			precisely that of the classical BPST instanton in Yang-Mills theory~\cite{Belavin:1975fg, tHooft:1976snw}!

		\subsection{Light and Massless Fermions} 

		As is familiar from standard instanton physics, the presence of light, charged fermions can dramatically alter a theory's $\theta$-dependence. In particular, any dependence on $\theta$ should vanish as we take any charged fermion's mass to zero and thus restore a chiral symmetry. 

		Such light fermions similarly affect dyonic physics. In the presence of a fermion of mass $m \ll m_\labM$,  the dyon's electric charge will no longer be localized to its core~\cite{Callan:1982ah,Callan:1982au} but will rather  be dispersed in the fermionic vacuum on a length scale of order $m^{\sminus 1}$. As $m \to 0$, this cloud grows to encompass all of space and an observer at finite distance would measure vanishing electric charge: this cloud screens the charge induced by the Witten effect. Furthermore, there exists a collection of fermionic excitations about the dyon that must be accounted for when computing our monopole loops.

		While a full analysis of this effect---and the  inclusion of multiple light fermions---is reserved for future work, we can easily understand how it impacts the dyon mass spectrum on dimensional grounds. Since the fermion dilutes the induced electric charge over a region roughly the size of its Compton wavelength, we expect that $r_* \sim m^{-1}$ in the estimate (\ref{eq:matching}), and so the dyonic mass spacing becomes of order $m_\Delta^2 \sim m_\labM m$. Since this spacing vanishes as $m \to 0$, so does the $\theta$-dependence of the dyon tower. 

		\subsection{Higher-Derivative Corrections and Validity }
		
		Our calculation assumed the dyon mass spectrum presented in~\eqref{eq:dyonmass}, which we expect to receive corrections in effective field theory when monopoles are not BPS. We should check that our result is robust against such corrections. These corrections can arise from higher derivative operators in the bulk effective Lagrangian, like $\left(F_{\mu \nu} F^{\mu \nu}\right)^2$ or $\big(F_{\mu \nu} \tilde{F}^{\mu \nu}\big)^2$, or higher powers of $(\partial_\mu \sigma + k A_\mu)$ in the worldline Lagrangian. These are related: the former add $\mb{B}^4, (\mb{E} \cdot \mb{B})^2, \mb{B}^2 \mb{E}^2$ and $\mb{E}^4$ terms to the energy density~$\rho$.
		Integrating $\rho$ outside the monopole core, similarly to the logic that led us to~\eqref{eq:matching}, implies that these terms modify the dyon mass spectrum. A series of terms of the form $c_{2j} \mb{E}^{2j}/\Lambda^{4(j-1)}$ in $\rho$ 
		generates corrections to the mass spectrum in even powers of $(n - \theta/{2\pi})$:
		\begin{equation} \label{eq:dyonhigherderiv}
		m_n^2 = m_\labM^2 + m_\labD^2 \sum_{j=1}^\infty \lambda_{2j}\!\left(n - \frac{\theta}{2\pi}\right)^{2j},
		\end{equation}
		where 
		$\lambda_{2j} \sim c_{2j} \left[e^2 k^2/\left(16\pi^2 (r_* \Lambda)^4\right)\right]^{j-1}$ is small when ${j > 1}$ (and $\lambda_1 \equiv 1$, by the definition of $m_\labD^2$). Terms involving powers of both $\mb{B}$ and $\mb{E}$ give subleading shifts to the definitions of $m_\labM^2, m_\labD^2$, and the $\lambda_j$.

\begin{figure*}[t]
\centering
\includegraphics[]{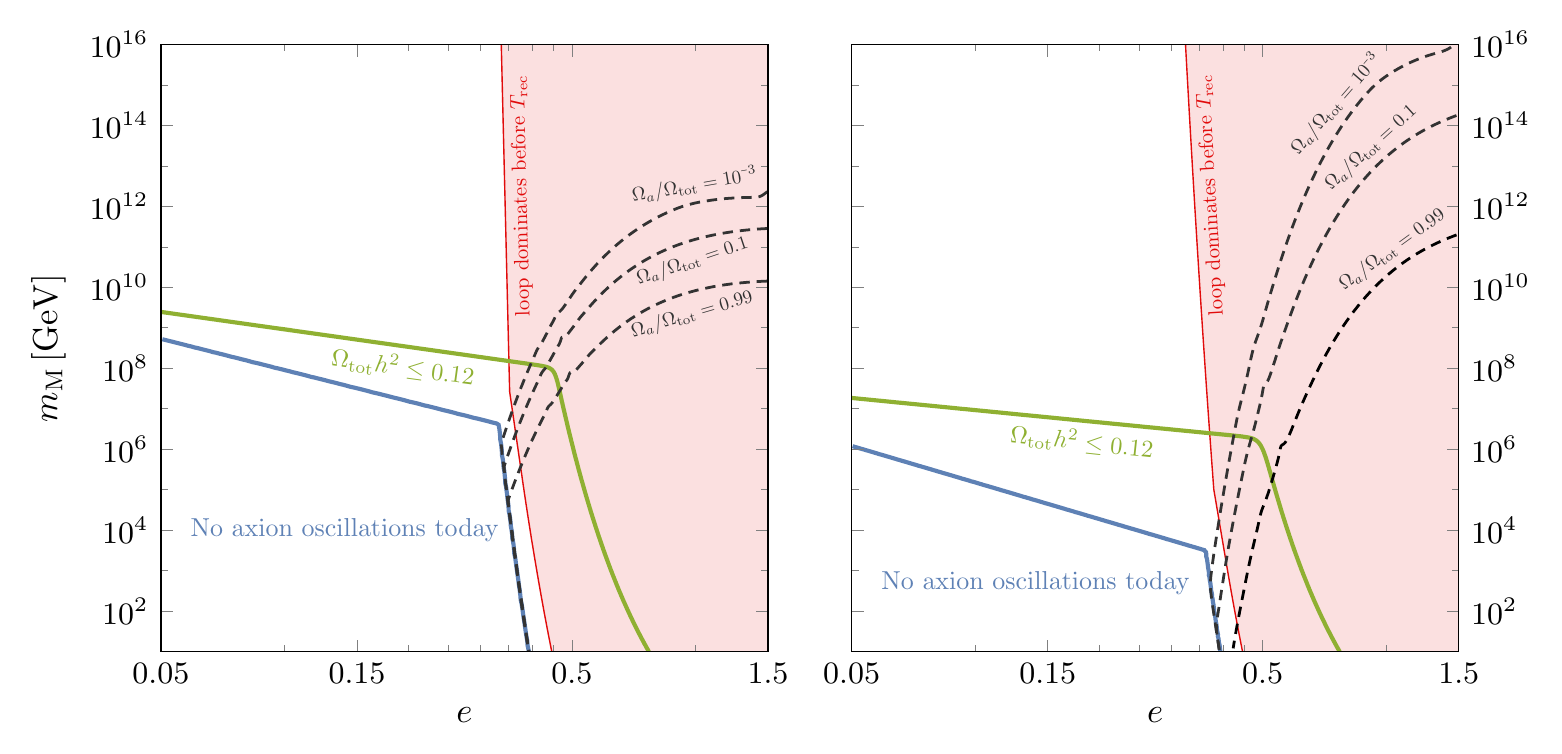}
\caption{Pink regions: axion oscillates with the monopole loop-induced mass dominating over the monopole background-induced mass before CMB formation at temperature $T_{\rm rec}$. Green solid line: the sum of axion abundance ($\Omega_a h^2$) and monopole abundance ($\Omega_\labM h^2$) today is $\Omega_{\rm tot}h^2$ = 0.12, above which the abundance overcloses the Universe. Black dashed lines indicate the fraction of axion dark matter in the total abundance of axion and monopole today. Below the blue line, the axion's mass is so small that it never oscillates. Left panel: monopole yield saturates the Kibble bound. Right panel: monopole yield is from a second order phase transition with a critical exponent $\nu = 0.5$. We fix $m_\labM r_c = {\pi}/{e^2}$, the critical temperature to be $T_c = 1/r_c$, and $f = 10^{15}$ GeV. }
\label{fig:parspace}
\end{figure*}%
		
		Repeating our earlier logic, we can sum the loop corrections~\eqref{eqn:eff_pot} using the mass spectrum~\eqref{eq:dyonhigherderiv}. Poisson resummation and relabeling $n - \theta/2\pi \to n$ then gives
		\begin{align}  \label{eq:Zlsaddle}
		V_\lab{eff}(\theta) &= -\sum_{\ell \in \mathbb{Z}} \int_{0}^{\infty}\!\frac{\ud \tau}{4 \tau} \frac{1}{(2\pi \tau)^{2}}\, \lab{e}^{-\frac{1}{2}m_\labM^2 \tau + i \ell \theta} \mathcal{Z}(\ell, \tau), \nonumber \\
	       \mathcal{Z}(\ell, \tau) &\equiv \int_{\sminus \infty}^{\infty}\!\ud n\, \lab{e}^{-2 \pi i n \ell - \frac{1}{2} m_{\scalebox{0.5}{$\Delta$}}^2 \tau(n^2 + \lambda_4 n^4 + \cdots)} 
		\end{align}		
		To evaluate the integral over $n$, we work in a saddle point approximation: defining $S_\ell(n)$ to be the function inside the exponent in~\eqref{eq:Zlsaddle}, we ask that $dS_\ell(n_*)/dn = 0$. Treating $\lambda_4$ as a perturbation, we find that
		\begin{equation}\label{eq:nsaddle}
		n_* = -\frac{2\pi i \ell}{m_\labD^2 \tau} - 2i \lambda_4 \left(\frac{2\pi \ell}{m_\labD^2 \tau}\right)^3 + \mathcal{O}(\lambda_4^2, \lambda_6).
		\end{equation}
		We require a small correction to the subsequent integral over $\tau$, dominated by the saddle at $\tau_* = 2\pi \ell/(m_\labM m_\labD)$. In particular, $m_\labD^2 \tau_* \ll 1$ (for small $\ell$), which calls for caution: the semiclassical approximation requires sufficiently small $\lambda_{2j}$. In particular,~\eqref{eq:nsaddle} implies that a small correction to $n_*$ at the saddle $\tau_*$ requires
		\begin{equation}
		|\lambda_4| \ll \frac{1}{2}\left(\frac{m_\labD^2 \tau_*}{2\pi \ell}\right)^2 \sim \frac{1}{2} \left(\frac{m_\labD}{m_\labM}\right)^2 \lesssim \frac{e^4 k^2}{8\pi^2} \frac{r_c}{r_*},
		\end{equation}
		where the last inequality can be derived from~\eqref{eq:matching} and the surrounding discussion. Along similar lines, we require $|\lambda_{2j}| \ll (m_\labD/m_\labM)^{2(j-1)}$. Focusing only on power counting in $e$, this requires that $|c_{2j}| \lesssim e^{2(j-1)}$. This will always hold when the operator $\mb{E}^{2j}$ is generated through loops of charged particles, as in the Euler-Heisenberg Lagrangian, where $|c_{2j}| \sim e^{2j}/(16 \pi^2)$. In the case with $r_* = r_c$, this verifies that our semiclassical calculation can be performed within the context of a sensible effective field theory in which higher derivative operators produce controllably small corrections. The case $r_* = r_0$ requires somewhat more care regarding the allowed range of the scale $\Lambda$, which we will not delve into here.

	 	\section{Phenomenological Applications}
		
There could potentially be many interesting phenomenological implications of this new monopole contribution to the axion potential, which calls for future work. Here we will only consider a hidden sector with a gauged $U(1)_d$ symmetry and a gauge coupling $e$ as a minimal example to show that this new contribution could play an important role in the cosmological evolution of axion-like particles. In the hidden sector, there is also a global $U(1)_{\textsc{pq}}$ symmetry which is broken spontaneously at the scale $f$ and results in a Goldstone boson, the axion $a$. 

In the presence of monopoles carrying magnetic charge under the $U(1)_d$, the axion obtains a mass from both the temperature-independent monopole loop, which is the new finding of our paper, and the temperature-dependent contribution from a monopole background as discussed in~\cite{Fischler:1983sc, Kawasaki:2015lpf, Nomura:2015xil,Kawasaki:2017xwt}. The monopole background could be generated via the Kibble-Zurek mechanism in a phase transition happening at a critical temperature $T_c$ in the early Universe~\cite{Kibble:1976sj, Zurek:1985qw}. The monopole yield satisfies the Kibble lower bound~\cite{Kibble:1976sj} and could be significantly above the bound if the phase transition is second order~\cite{Zurek:1985qw, Murayama:2009nj}. In this model, both axions and monopoles could be components of dark matter. The relevant parameter space is shown in Fig.~\ref{fig:parspace}, assuming that the visible and dark sectors share a common temperature at early times. We see that when the gauge coupling $e$ is large, $e \gtrsim 0.5$, the monopole loop-induced axion mass would dominate over the contribution from the monopole background before the CMB formation. In addition,  the axion abundance is negligible when $e \lesssim 0.5$, but it could take over that of the monopoles at larger values $e \gtrsim 0.5$. In order not to overclose the Universe, the monopoles must not be very heavy~\cite{Zeldovich:1978wj,Preskill:1979zi}. Fig.~\ref{fig:parspace} establishes that the new effect we discuss can modify the cosmology of axions and monopoles; it would be interesting to incorporate it in a wider range of models in the future.

\section{Conclusions}
		
In this article, we have presented and computed a new contribution to the vacuum axion potential from magnetic monopole loops, when the axion is coupled to an abelian gauge field. Much more remains to be studied, both in developing the formalism and exploring the phenomenological and cosmological implications. We briefly comment on some possible future directions: {\it 1)} We have assumed $V_\lab{eff}(\theta)$ is dominated by a single monopole loop, but there are long-range Coulomb interactions between the monopoles. Their effect on the semiclassical expansion should be explored. {\it 2)} We found that the action of the monopole-loop instanton in the critical 't~Hooft-Polyakov case is that of a BPST instanton, $8\pi^2/\wG^2$. This may be a harbinger of a stronger statement: we expect that the monopole-loop instantons can be continuously deformed into nonabelian instantons. If not, the theory would have an unbroken global $(-1)$-form $U(1)$ symmetry in 4d, and a $(d-5)$-form symmetry in higher dimensions~\cite{Heidenreich:2020pkc}. Similarly, in cases where $U(1)$ gauge fields arise on D($n+3$)-branes wrapped on $n$-cycles in extra dimensions, one obtains axion potentials either from wrapped Euclidean D$(n\!-\!1)$-branes or from magnetic monopoles, which are D($n+1$)-branes ending on the D$(n+3)$-branes. 
The
winding of $\sigma$ on the monopole worldline in 4d arises from a nonvanishing field strength on the D$(n+3)$-brane, which via the worldvolume  Chern-Simons term, is equivalent to D($n\!-\!1$)-brane charge. Again, we expect that the monopole-loop instanton can be continuously deformed into a Euclidean brane instanton in this context. These deformations between instantons should be constructed more explicitly. {\it 3)} We demonstrated that this new contribution could be important in a hidden sector model with the axion coupling to a dark $U(1)_d$. Consider an axion coupling to the standard model photon instead. Does this imply a minimum mass of the axion, even without nonabelian instantons? What are the effects of multiple fermions, present in the standard model? Future work answering these questions will directly connect the effect we have presented with ongoing experiments.

		\section*{Acknowledgments}
		
MR especially thanks Jake McNamara for related discussions during a previous collaboration. We thank Prateek Agrawal, Jake McNamara, and Ofri Telem for useful discussions. JF is supported in part by the DOE grant DE-SC-0010010 and the NASA grant 80NSSC18K1010. KF is supported by the National Science Foundation Graduate Research Fellowship Program under Grant No.~DGE1745303. MR and JS are supported in part by the NASA Grant 80NSSC20K0506. MR is supported in part by the DOE Grant DE-SC0013607 and the Alfred P.~Sloan Foundation Grant No.~G-2019-12504.

		\bibliography{monopot}

\end{document}